# Finding Signs of Life in Transits: High-Resolution Transmission Spectra of Earth-like Planets around FGKM Host stars

Lisa Kaltenegger[1,2] & Zifan Lin[1,2,3]
[1]Cornell University, Astronomy, and Space Sciences Building, Ithaca, NY 14850, USA
[2]Carl Sagan Institute, Space Science Building 311, Ithaca, NY 14850, USA
[3]Department of Earth, Atmospheric and Planetary Sciences, MIT, Cambridge, MA 02139, USA

## Abstract:

Thousands of transiting exoplanets have already been detected orbiting a wide range of host stars, including the first planets that could potentially be similar to Earth. The upcoming Extremely Large Telescopes and the James Webb Space Telescope will enable the first searches for signatures of life in transiting exoplanet atmospheres.

Here, we quantify the strength of spectral features in transit that could indicate a biosphere similar to the modern Earth on exoplanets orbiting a wide grid of host stars (F0 to M8) with effective temperatures between 2,500 and 7,000K: transit depths vary between about 6,000ppm (M8 host) to 30 ppm (F0 host) due to the different sizes of the host stars.

$CO_2$ possess the strongest spectral features in transit between 0.4μm and 20μm. The atmospheric biosignature pairs $O_2+CH_4$ and $O_3+CH_4$ – which identify Earth as a living planet – are most prominent for Sun-like and cooler host stars in transit spectra of modern Earth analogs. Assessing biosignatures and water on such planets orbiting hotter stars than the Sun will be extremely challenging even for high resolution observations.

All high-resolution transit spectra and model profiles are available online: They provide a tool for observers to prioritize exoplanets for transmission spectroscopy, test atmospheric retrieval algorithms, and optimize observing strategies to find life in the cosmos.

In the search for life in the cosmos, transiting planets provide the first opportunity to discover whether or not we are alone, with this database as one of the keys to optimize the search strategies.

**Keywords:** Transits, Biosignatures, Extrasolar rocky planets, Habitable planets, Exoplanets, Observational astronomy, Transmission spectroscopy, High resolution spectroscopy, Exoplanet atmospheres, Exoplanet atmospheric composition, Astrobiology

## Introduction:

More than 4,000 known exoplanets have been identified to date (exoplanets.nasa.gov Oct 2020) including the first potentially Earth-like planets (see e.g., Kaltenegger & Sasselov 2011, Kane et al.2016, Johns et al.2018, Berger et al.2019). These planets orbit a wide range of host stars.

Sagan et al.(1993) analyzed the spectrum of Earth - observed by the Galileo probe - searching for signatures of life. The study concluded that the large amount of $O_2$ and the simultaneous presence of $CH_4$ traces are strongly suggestive of biology (see Lovelock 1965, Lederberg 1965). A study by the authors (Kaltenegger et al.2020a) found that the biosignature pairs $O_2+CH_4$ and $O_3+CH_4$ could identify a transiting Earth as a living planet from about 1 to 2 billion years ago.

Several studies discuss a wide range of atmospheric biosignatures for conditions differing from modern Earth e.g. for abiotic conditions (e.g. Kaltenegger et al.2007, Kasting et al.2014, Lyons et al.2014, Arney et al.2016, Rimmer & Rugheimer 2019, Kaltenegger et al.2020b) and how observations could possibly discriminate the different evolutional paths of a planet (e.g. Lincowski et al.2018), which is beyond the scope of this study. Here we focus on transit spectra of modern Earth atmospheres with similar outgassing rates, surface pressure and stellar irradiation with a tested 1D model (see Madden & Kaltenegger 20202a): This model uses wavelength-dependent surface albedos that distinguishes the influence of surfaces and clouds discussed in detail in Madden & Kaltenegger (2020a). Here we provide a database of transit spectra that spans planets orbiting F0 to M8 host stars to assess the variability in spectral transit feature strength for exoplanets, based on their host star. This database provides a tool for observers that covers the full search space of stellar hosts for Earth-like planets (model and high-resolution spectral are available online, add link here when accepted).

The stellar energy distribution (SED) of the host star influences the chemical composition of a planet's atmosphere (see e.g., Kasting et.al.1993, Segura et.al.2005, Hedelt et.al.2013, Rugheimer et.al.2015, Ranjan&Sasselov2016, Rugheimer & Kaltenegger 2018, Kaltenegger 2017, Fuji et al.2018, Wunderlich et al.2019, Kaltenegger et al.2020b). In addition to altering photochemistry - especially for tidally locked planets around cool stars (see Chen et al 2019) - changes in the stellar SED can influence atmospheric circulation and climate strongly (see. e.g., Pierrehumbert 1995, Leconte et al.2013, Wolf & Toon 2015, Koll & Abbot 2015, Kopparapu et al.2017, Fauchez et al.2019, Chen et al.2019, Suissa et al.2020). Planetary rotation rates can also impact the dynamics of Earth-like planets (see e.g., Yan et al.2013, Wolf & Toon 2013, Kopparapu et al.2017, Haqq-Misra et al.2018, Way et al.2018, Komacek & Abbot 2019). Recent results (Chen et al.2019) found that the stratospheric day-to-nightside mixing ratio differences even on tidally locked planets remain low (<20%) across the majority of biosignature gases, including the biosignature pairs considered in this paper, which is encouraging. 3D models should provide guidance for 1D models, which can assess a wider parameter space. For now, we have to leave it to future studies to compare sets of planets using 1D and 3D GCM models with similar stellar input, planetary parameters, cloud coverage and photochemistry schemes to produce a more accurate guidance.

To assess the effect of the host star's irradiation on Earth-like atmospheres for a wide grid of host stars the authors previously modeled modern Earth-like planets around 12 FGK stars (see details in Madden & Kaltenegger 2020a, Rugheimer et al.2013) and 10 M star (Rugheimer et al.2015) ($T_{eff}$ 7,000–2,500 K), using a 1D climate-photochemistry-Radiative Transfer model with i) a constant averaged surface-cloud albedo (ExoPrime: Rugheimer et al.2013, 2015) and ii) a wavelength-dependent surface-cloud albedo (ExoPrime2: Madden & Kaltenegger 2020a). For these models, the authors provided online, high-resolution templates as a tool to optimize observations of directly imaged as well as exoplanets seen just before and after secondary eclipse as "Pale Blue Dots" (Madden & Kaltenegger 2020b): A remote observer could identify a biosphere for such planets in high-resolution emission and reflection spectra for all modeled host stars, with signature strength decreasing in reflected light for cooler host stars.

However, transit spectroscopy is the most commonly used methods for probing exoplanet atmospheres to date: Upcoming ground-based Extremely Large telescopes (ELTs) and space-based missions like JWST plan to characterize the atmosphere of transiting planets and search for signs of life in their atmospheres (see e.g, Gialluca et al. 2021, Wunderlich et al. 2019). Thus, in this paper we model the transit spectra of the planet models discussed in Madden & Kaltenegger (2020a) to assess the changes of the strength of spectral features in transit of Earth-like planets for a wide range of host stars.

Upcoming spectrographs like HIRES (0.3-2.5μm) and METIS (3-19μm) on the ELT are designed for a resolution of R≈100,000 (Ramsay et al.2020) to distinguish Earth's atmospheric features from potentially habitable exoplanets' (e.g. Snellen et al.2015, Fischer et al.2016, Lopez-Morales et al.2019, Serindag & Snellen 2019, Lin & Kaltenegger 2020). Therefore, we provide a database of transit spectra with 0.01cm$^{-1}$ steps size - a minimum resolution (R=λ/Δλ) of R ≥ 100,000 from 0.4 to 10μm and R ≥ 50,000 from 10 to 20μm - as a tool for observers to optimize observations with upcoming telescopes, train retrieval algorithms and prioritize targets.

Which spectral features could indicate a nominal biosphere similar to modern Earth's on exoplanets that transit different host stars?

## 2. Methods:

This spectral library is based on self-consistent simulated exoplanet atmospheres in 1D with wavelength dependent surface and cloud reflectivity (as described in Madden & Kaltenegger 2020a). We used the same approach as discussed in Madden & Kaltenegger (2020a) to include 3 new models for M star planets in this paper. As described in methods and in detail in Madden & Kaltenegger (2020a) the incident stellar flux at the planet's location, $S_{pl}$, decreases with stellar temperature to provide similar planetary surface temperatures (table 1). All host star spectra are shown in Fig.1, temperature/pressure as well as major chemical mixing ratios are shown in Fig.2. The corresponding transmission spectra are shown in fig.3 and fig 4 in low resolution (≈700) and in fig.5 in high resolution (≈100,000) for the strongest biosignature features.

### 2.1 Models Description and numerical setup

Here, we define `Earth-like' to refer to a one Earth-radius, one Earth-mass planet with similar incident irradiation, outgassing rates, surface pressure, composition, and cloud coverage to modern Earth (see details in Madden & Kaltenegger 2020a). Modeling similar planets allows us to evaluate the effect of the host star on the transit spectra of Earth-like planets.

Madden & Kaltenegger (2020a) explored the effect of different surfaces on the climate of planets orbiting host stars from F0 to K7 in 250K stellar surface temperature ($T_{eff}$) steps with Exo_Prime2, which uses wavelength-dependent surface albedos that distinguishes the influence of surfaces and clouds to assess the effect of a planet's surfaces on the climate. Similar models for M-star planets described in Rugheimer et al.(2015), did not take the wavelength dependence of the surface albedo into account. Thus, in this paper we update the nominal M-star models (Rugheimer et al.2015) using Exo-Prime2 for three specific M stars with updated UV observations: ADLeo, Proxima Centauri, and TRAPPIST-1 (see details for stellar models in O'Malley-James & Kaltenegger 2019) to expand our model grid to M-stars with well-observed spectra. All planetary models for F to K stars are described in Madden & Kaltenegger 2020a. Stellar models for F0 to K7 hosts are described in Rugheimer et al.(2013) and models for M hosts are described in O'Malley-James & Kaltenegger (2019) in detail (summarized in table 1 and fig.1).

The atmospheric models assume modern Earth's surface composition: 70% ocean (albedos of seawater) and 30% land (a combination of basalt, granite, sand, trees, grass, and snow) (following Kaltenegger et.al.2007). Surface albedos for modeling the atmosphere were taken from the USGS and ASTER spectral libraries (Baldridge et.al.2009, Kokalyet et.al.2017, Clark et.al.2007).

Similar stellar incident - top of the atmosphere - irradiation would increase the surface temperature of planets orbiting cooler stars considerably because of the reduced effectiveness of Rayleigh scattering at longer wavelengths and the increase in near-IR absorption by $H_2O$ and $CO_2$ as the star's spectral peak shifts to longer wavelengths (see e.g., Kasting et al.1993). Thus, planet models around cooler stars receive lower total stellar incident flux in our models to generate surface temperatures similar to modern Earth across star types (discussed in detail in Madden & Kaltenegger 2020a). Temperature profiles and mixing ratios for the major gases in the atmosphere are shown in figure 2 (see also Madden & Kaltenegger 2020a).

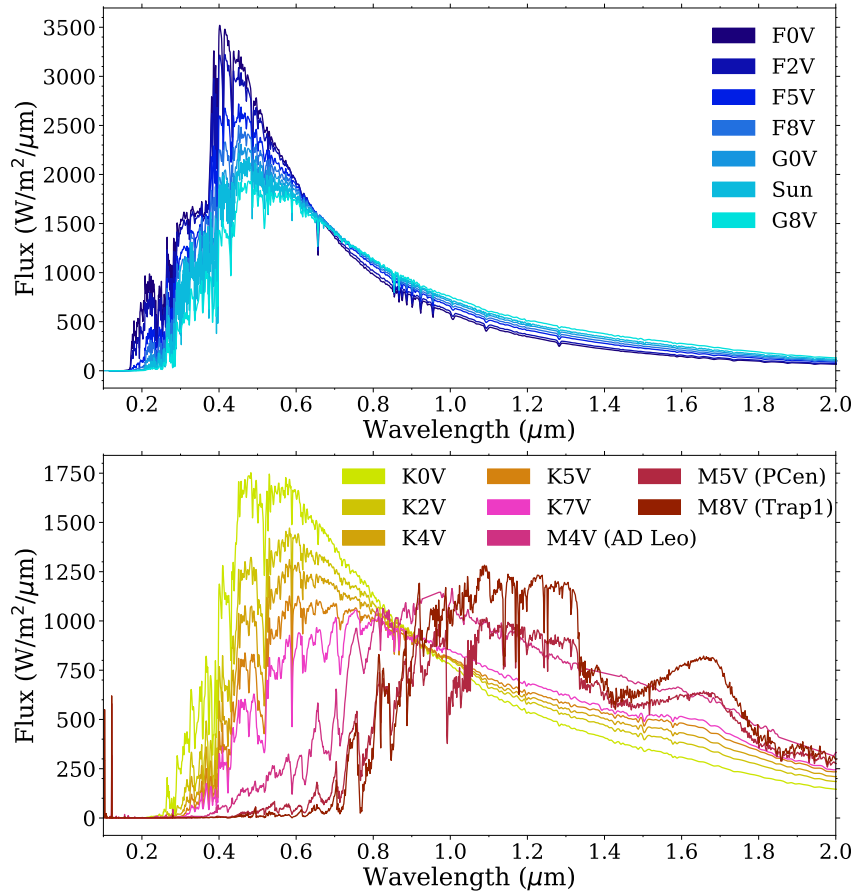

**Figure 1**: Composite stellar input spectra from IUE or MUSCLE observations merged to ATLAS photosphere model at 0.3μm for each grid star – (top) F and G and (bottom) K and M host star – shown at a resolution of 700 for clarity. Stellar files extend to 20μm but are only shown until 2μm for clarity (for stellar model details see Rugheimer et al.2013 and O'Malley-James & Kaltenegger 2019).

The average surface temperature for the planet models is 288K +/- 2% for all models, with the model orbiting the Sun at 288K. Ozone column depth decreases with decreasing UV environments with a minimum around K7 hosts, that provides the lowest UV environment of our grid stars (see table 1 and fig.2). $CH_4$ levels increase for lower UV environments (see also e.g., Segura et al.2005, Rugheimer et al.2015). $N_2O$ is destroyed by UV radiation in the upper atmosphere, thus its concentration also increases for lower UV environments with a maximum around K7 and decreases again for the active M stars as expected (see also e.g., Grenfell 2017, Rugheimer & Kaltenegger 2018). Note that our models focus on time-averaged stellar spectra. Time-evolving space weather events might lead to a temporary reduction of $CH_4$ due to the interaction of charged particles derived from flares and stellar winds with a planet's atmosphere (see e.g., Segura et al. 2010, Tilley et al. 2019, Chen et al. 2020, review: Airapetian et al.2020).

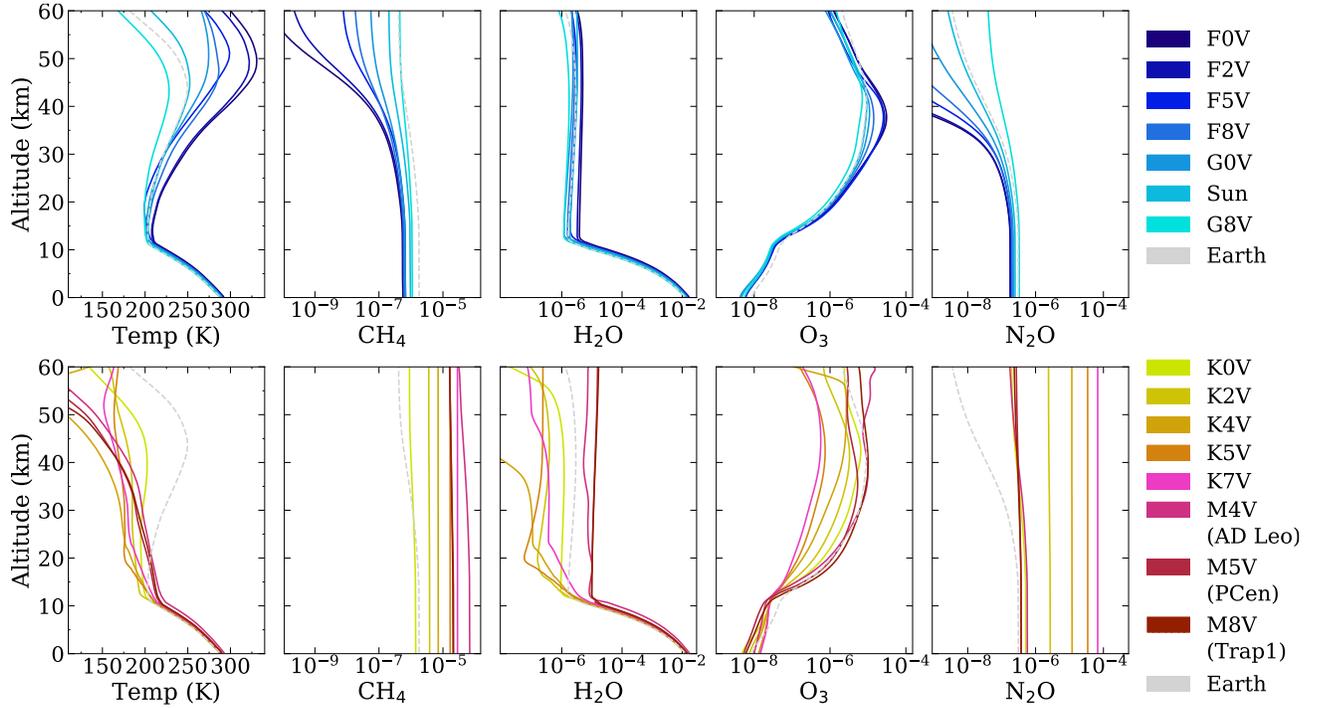

**Figure 2:** Temperature profile and mixing ratios for the major atmospheric gases for Earth-like planets around (top) F and G and (bottom) K and M grid host stars, with modern Earth radius, mass, pressure, and outgassing ratios (see Madden & Kaltenegger 2020a).

**Table 1**: List of host star models using representative IUE stars (F to K stars) and MUSCLEs UV data (M stars). The table gives the name and measured $T_{eff}$, the $T_{eff}$ which corresponds to our grid of stars, approximate stellar type, surface temperature, incident flux at the planet's orbit in Earth irradiation ($S_{pl}/S_0$) and $O_3$ column depth for a modern Earth-like planet model orbiting this grid of host stars (details on stars in Rugheimer et al.2013 and O'Malley-James & Kaltenegger 2019).

| Star | | | | | Model Planet | | |
|---|---|---|---|---|---|---|---|
| Name | $T_{eff}$ (K) | Spectral Type | $T_{eff}$ (K) Grid | $S_{pl}/S_0$ | Surface Temp (K) | Ozone Column Depth (cm$^{-2}$) | Transit Depth surface (ppm) |
| η Lep | 7060 | F0V | 7000 | 1.11 | 291.73 | 9.7813E+18 | 31.9 |
| σ Boo | 6730 | F2V | 6750 | 1.09 | 290.48 | 9.2306E+18 | 38.3 |
| π³ Ori | 6450 | F5V | 6500 | 1.06 | 290.10 | 8.6086E+18 | 42.8 |
| ι Psc | 6240 | F7V | 6250 | 1.04 | 289.65 | 6.3618E+18 | 58.3 |
| β Com | 5960 | F9V/G0V | 6000 | 1.02 | 289.15 | 5.3049E+18 | 66.9 |
| Sun | 5778 | G2V | 5750 | 1.00 | 287.91 | 4.9713E+18 | 83.9 |
| τ Ceti | 5500 | G8V | 5500 | 0.98 | 286.33 | 3.5468E+18 | 101.3 |
| HD 10780 | 5260 | K0V | 5250 | 0.96 | 287.04 | 2.7420E+18 | 121.8 |
| ε Eri | 5090 | K2V | 5000 | 0.92 | 286.49 | 1.6291E+18 | 149.2 |
| ε Indi | 4730 | K4V | 4750 | 0.91 | 288.03 | 1.0300E+18 | 178.8 |
| 61 Cyg A | 4500 | K5V | 4500 | 0.89 | 287.45 | 8.0789E+17 | 204.9 |
| BY Dra | 4200 | K7V | 4250 | 0.88 | 288.94 | 7.3339E+17 | 287.8 |
| AD Leo | 3400 | M4V | 3500 | 0.86 | 292.15 | 5.1721E+18 | 551.7 |
| Proxima Cen | 3050 | M5V | 3000 | 0.84 | 289.04 | 4.1668E+18 | 3529 |
| TRAPPIST-1 | 2559 | M8V | 2500 | 0.82 | 289.59 | 6.1798E+18 | 6130 |

Earlier research by several teams shows the dependence of snowball-deglaciation loop on the stellar type, which can have a strong impact on long-term sustained surface habitability (e.g., Abe et al.2011, Shields et al.2014, Abbot et al.2018). For planet models with different surface compositions (see e.g., Abe et al.2011, Shields et al.2014, Abbot et al.2018, Madden & Kaltenegger 2020a), the authors showed that surfaces with highly varying reflection from the visible to near-IR can influence the surface temperature across star types considerably if their surface coverage is high. The high-resolution reflection and emission spectra for these models of Earth-like planets with different surfaces are provided online in Madden & Kaltenegger (2020b). The corresponding transit spectra for Earth-like planets are presented in this paper.

The transit depth shown in table 1 is between about 6,000 ppm and 30 ppm – using only the planet's surface for its radius in this simplified calculation. Spectral features further increase this transit depth because they increase the size of the planet at the wavelength they absorb at.

## 2.2. Generating transmission spectra of Earth-like planets

We model the high-resolution transmission spectra for nominal Earth-like planets from 0.4 to 20µm at a resolution of $0.01 cm^{-1}$ using Exo-Prime2, a 1D iterative climate-photochemistry code coupled to a line-by-line- radiative transfer code (details in Madden & Kaltenegger 2020a). The Radiative Transfer component was developed for stratospheric retrieval on Earth, adapted to rocky exoplanets (see Traub&Stier1976, Traub&Jucks2002, Kaltenegger et al.2007, Kaltenegger & Traub 2009) and validated from the visible to infrared through comparison to Earth seen as an exoplanet by missions like the Mars Global Surveyor, EPOXI, multiple Earthshine observations and Shuttle data (Kaltenegger et al.2007, Kaltenegger & Traub 2009, Rugheimer et al.2013).

We include the most spectroscopically relevant molecules in our calculations: $C_2H_6$, $CH_4$, $CO$, $CO_2$, $H_2CO$, $H_2O$, $H_2O_2$, $H_2S$, $HNO_3$, $HO_2$, $N_2O$, $N_2O_5$, $NO_2$, $O_2$, $O_3$. $OCS$, $OH$, $SO_2$, using the HITRAN2016 line lists (Gordon et al.2017) as well as Rayleigh scattering. We divide the planet atmosphere into 52 layers: for each atmospheric layer line shapes and widths are calculated individually with Doppler- and pressure-broadening with several points per line width.

Deeper atmospheric regions can deflect light away from a distant observer (see e.g., Sidis & Sari 2010, Garcia Munoz et al.2012, Betremieux & Kaltenegger 2014, Robinson 2017), which sets the lowest level Earth can be probed to in primary transit to about 13km. For Earth, that does not affect the transit spectrum significantly because Earth can only be probed down to about 13km in the UV to IR wavelength range because of absorption and refraction in the atmosphere (see e.g., Kaltenegger & Traub 2009). For other host stars the depth to which the atmosphere can be probed down, due to refraction only, varies between 15.7km and 0km: 15.7km(F0V), 13.8km(F7V), 12.6km(Sun), 11.7km(G8V), 9.6km(K2V), 6.6km(K7V), 1.7km(M4V), 0km(M5V), 0km(M8V) (interpolated from results by Betremieux & Kaltenegger 2014).

Clouds close or on the terminator region - that is probed during primary transit - will obscure the spectral features in transmission below the cloud layer (e.g., Seager et al.2005, Kaltenegger & Traub 2009, Robinson et al.2011). The distribution of clouds for Earth-like planet models orbiting different host stars depend on parameters such as planetary rotation rate and are therefore uncertain because such information does not exist for exoplanets. Thus, we show the effect of a 100% opaque cloud layer at the terminator by adding a dashed line in our transmission spectra (fig.3) at 6km – basing this choice on the middle layer of Earth clouds, which are located at about 1km, 6km, and 12km. This hypothetical cloud layer at the terminator only affects the transit spectra for the M-star planets, which can be probed below that altitude. Higher opaque cloud layers will reduce the spectral features shown, lower cloud layers would increase the detectable spectral features in the transit spectra from the 6km cloud line shown in fig.3 (see e.g. Kaltenegger et al. 2007, Fauchez et al. 2019, Suissa et al. 2020, Komacek et al 2020).

Our database contains clear atmospheres at the terminator for all models. Clouds can easily be added to these by replacing the effective height of the planet in the transit spectra below the hypothetical cloud layer with the cloud layer height. We show the transit spectra binned to a resolution of $R = \lambda/\Delta\lambda = 700$ in fig.3 and fig.4 using a triangular smoothing kernel in the figures for clarity, corresponding to the resolving power for JWST instruments. However, all models and spectra can be downloaded at full resolution online (link here once accepted).

## 3. Results & Discussion

The strength of the spectral features of atmospheric chemicals in transit - including the biosignature pairs $O_2+CH_4$ and $O_3+CH_4$ - change depending on the host stars (F0 to M8). If a biosphere existed on nominal Earth-like planets orbiting these diverse host stars, which spectral features could upcoming telescopes like the ELTs and JWST search for?

### 3.1 How transit spectra features vary with host stars

The strongest spectral features - due to a gas blocking transmitted starlight - between 0.4 and 20μm are created by $CO_2$, $O_3$, $CH_4$, $H_2O$, and $O_2$ (labeled in fig.3 and fig.4). The strength of a spectral feature in the atmosphere is governed by the abundance of different chemicals as well as how deep a planet's atmosphere can be probed in transit, with cooler star planets being probed deeper. The grid of transit spectra models for Earth-like planets around a wide range of host stars from F0 to M8 (7,000K to 2,500K) is shown in fig.3 at a resolution of 700 for clarity.

Several strong $CO_2$ features at 2.6, 4.3, and 15μm show similar spectral feature strength for all modeled planets. The $CO_2$ feature at 4.3μm shows the largest absorption of the overall transmission spectrum between 0.4 to 20μm.

$H_2O$ features' strengths are similar between models, with the strongest features indicated in figure 3 at 0.95, 1.14, 1.4, 2.5, 6.5, and 17-25μm. However, due to their low effective altitude, these spectral lines get strongly diminished if the planet's atmosphere cannot be probed deeply for hotter host stars as shown in fig.3 and fig.4. Water features in the transmission spectra of an Earth-like planet orbiting hot F stars are extremely hard to discern in its transmission spectrum.

Oxygen feature's strength at 0.76μm is similar between models but is also strongly affected by the depth the atmosphere can be probed down to for hot host stars. Increased UV radiation – around hot F stars as well as active M stars - result in higher ozone concentration in the atmosphere of planet models as seen in the ozone feature at 9.6μm. While the strength of the spectral feature of the Chappuis ozone band in the visible – between 0.4 and 0.65μm - also increases with UV activity of the host star the effect is smaller and strongly influenced by the refraction in the atmosphere, especially for hotter host stars.

Methane concentration and features' strengths increase for cooler host stars, with the strongest $CH_4$ features in the modeled wavelength range at 7.6μm, followed by 3.3μm and 2.4μm. The concentration of $N_2O$ as well as its spectral feature strength at 3.9μm, 4.5μm, and 17μm increases with decreasing UV activity of the host star.

Figure 3 shows the corresponding transit depths for the transmission spectra for Earth-like planet models orbiting three of the grid host stars F0 (top), Sun (middle), and M8 (bottom). The locations of the most prominent spectral features are shown as labels. The y-axis shows effective height above the surface of the atmosphere of a planet on the left — the addition in size of the planet due to absorption and refraction in the planet's atmosphere — which is comparable between models, as well as the transit depth on the right, which increases by 2 orders of magnitude because of the decreasing size of the host star: about 6,000ppm (M8 host), 80ppm (Sun) and 30 ppm (F0 host) as shown in fig.3.

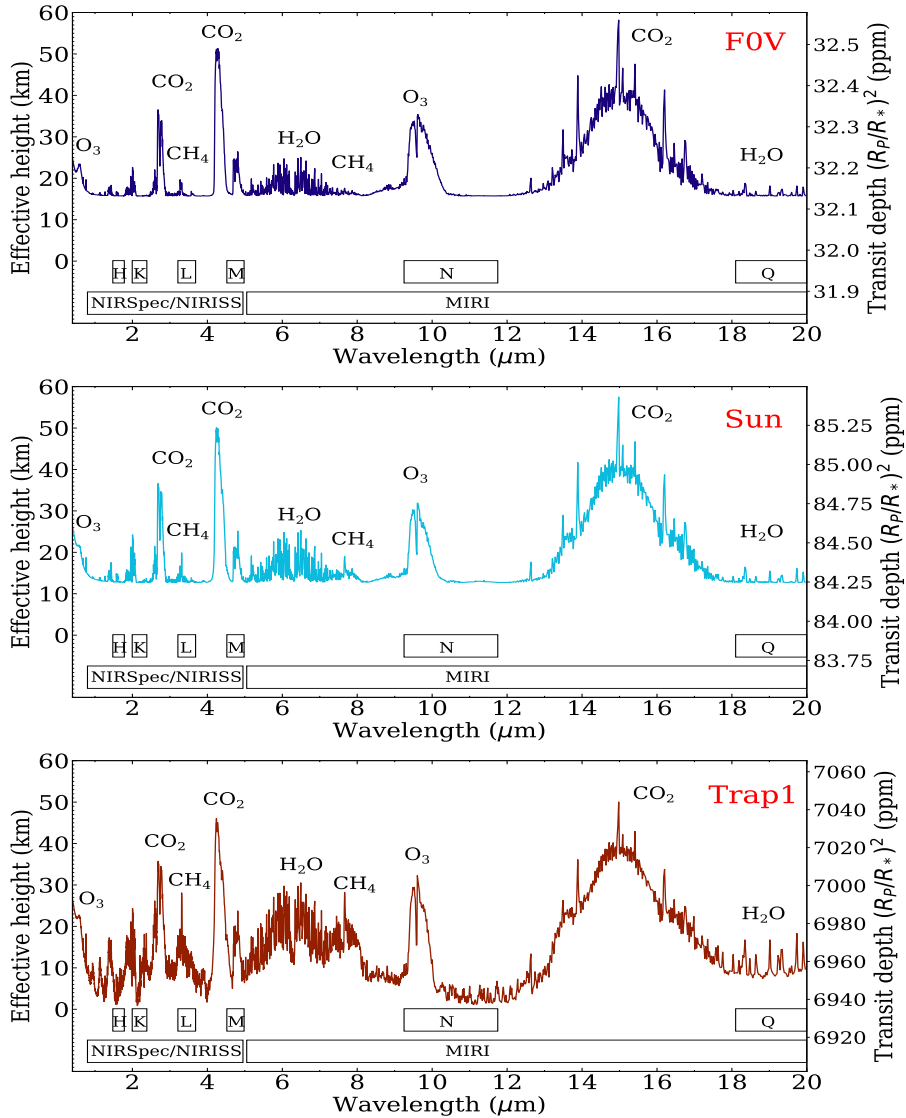

**Figure 3**: Transmission spectra (0.4-20µm) shown in effective height and transit depth for Earth-analog planet models orbiting three of the grid host stars F0 (top), Sun (middle), and M8 (bottom) hosts at a resolution of 700. Locations of the most prominent spectral features are shown as labels. The y-axis shows effective height of the atmosphere of a planet above the ground as well as (right) transit depth, which increases by 2 orders of magnitude because of the decreasing sizes of the host stars. *Data online at 10.5281/zenodo.4541545*

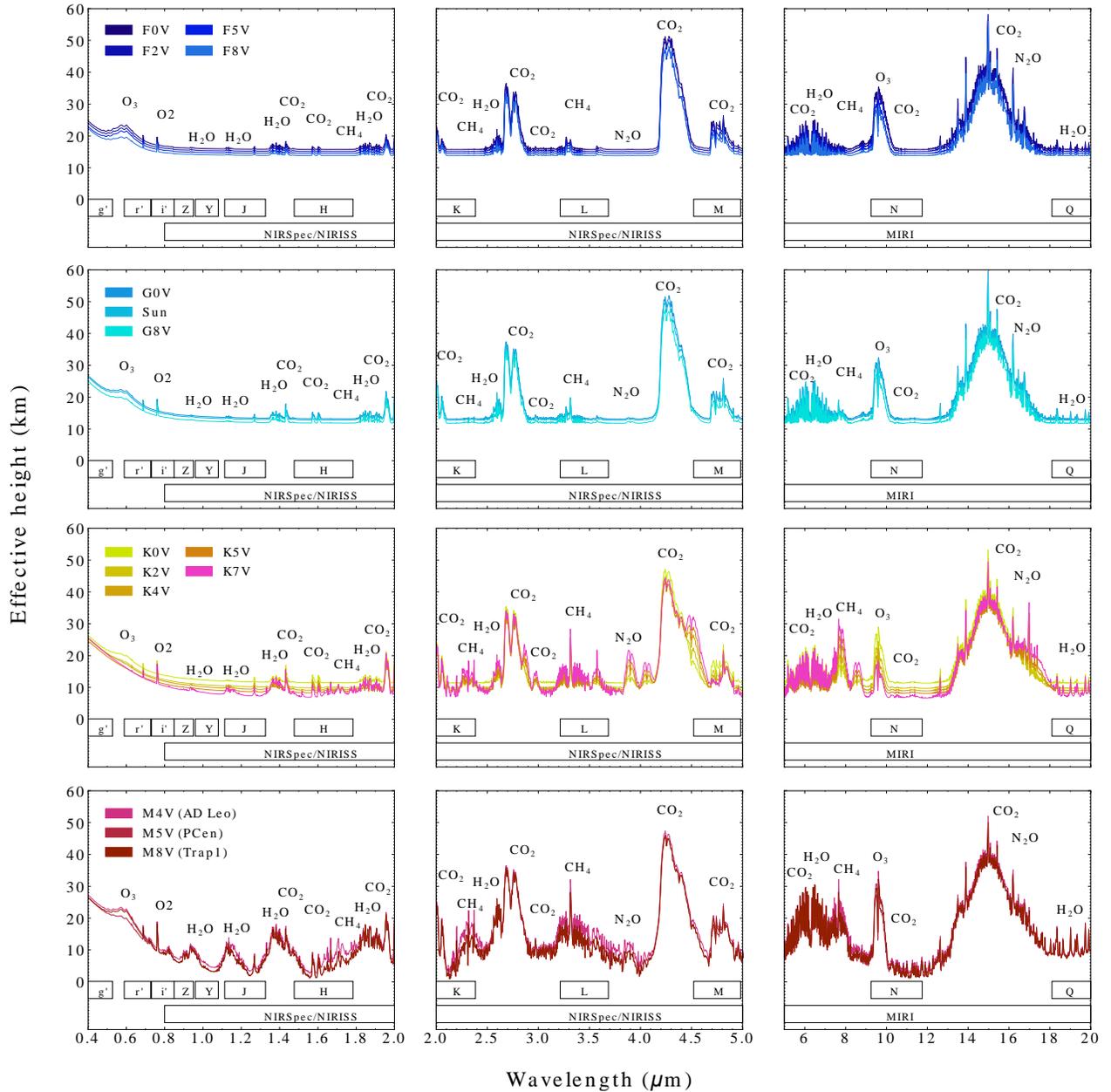

**Figure 4**: Transmission spectra shown as effective height of Earth-analog planets orbiting a wide range of host stars, shown at a resolution of 700 for Fstars (top), Gstar (2nd row), Kstar (3rd row) and Mstar (bottom row) hosts. The spectra are shown in three wavelength ranges for clarity – (left) VIS (0.4-2.0μm), (middle) NIR (2.0-5.0μm), and (right) IR (5-20μm). Locations of the most prominent spectral features are shown as labels. *Data online at 10.5281/zenodo.4541545*

Studies of synchronously rotating ocean-covered planets at the inner edge of the Habitable Zone of M stars find that such synchronous rotation can produce strong substellar convection that transports water to the upper atmosphere (e.g., Kopparapu et al.2017). It would thus increase the strength of the water features shown in fig.3 and fig.4 considerably.

## 3.2 How transit spectra biosignatures vary with host stars

Figure 5 shows the strongest spectral features for the biosignature pairs $O_2$+$CH_4$ in the visible to near-IR spectral range (0.4 to 3µm) and for $O_3$+$CH_4$ in the near-IR to IR range (3 to 20µm) in high resolution for 6 representative stellar hosts (F0,F2,Sun,K2,M5,M8).

The maximum effective height of the oxygen feature in high resolution ($\approx$100,000) is about twice its effective height in low ($\approx$700) resolution shown in fig.3 because of the narrow $O_2$ spectral feature at 0.76µm. The $O_2$ feature strength is similar for all models, decreasing slightly for cooler host stars. Ozone concentration as well as its features strength increases, however, with increasing UV flux from the star. This is reflected in the strongest $O_3$ feature strength at 9.6µm, which shows the strongest feature for hot F stars and active M stars. Combined with methane concentration and thus the strength of its spectral features increasing for cooler stars, the biosignature pairs $O_2$+$CH_4$ and $O_3$+$CH_4$ become increasingly difficult to discern for hotter stars than the Sun even in high-resolution and easier for cooler stars.

Even the strongest methane spectral features at 7.6µm, 3.3µm, and 2.4µm produce low effective height increases in the atmosphere of planets models orbiting hot host stars. These features are difficult to discern when the atmosphere cannot be probed to the ground due to refraction. For the Sun, the feature becomes discernable and increases in strength with decreasing host star temperature. Note that the 3.3µm is slightly stronger than the 2.4µm shown in figure 4. However, the 3.3µm feature requires a larger extension of wavelength coverage for visible concepts to include, therefore we show the 2.4µm feature in fig.5. Note that fig.5 shows clear atmosphere models: Opaque cloud layers would reduce the spectral features shown, depending on the cloud height.

## 4. Conclusions

In this paper we provide high-resolution transit spectra for a wide range of host stars (F0 to M8) as a tool for observers to plan upcoming observations for exoplanets. All the spectra and model profiles for Earth-like planets are available online (10.5281/zenodo.4541545):

This spectral database shows that signs of life – the biosignature pairs $O_2$+$CH_4$ and $O_3$+$CH_4$ on Sun-like and cooler host stars leave strong spectral features between 0.4 to 20µm in the transmission spectra of nominal Earth-like planets – if life like on Earth exists on such exoplanets. However, spectral signatures of water in transmission of Earth-like planets orbiting hot F stars are extremely hard to discern even in high resolution. Thus, younger Earth models, with higher methane would provide better targets for the search for life on transiting planets orbiting hot F stars.

This high-resolution spectral database provides a tool for observers to prioritize exoplanets to observe depending on the instrument capabilities, test retrieval algorithms, and interpret upcoming observations in our quest to find life in the universe.

**Acknowledgment**
This work was supported by the Carl Sagan Institute and the Brinson Foundation. The authors thank Jack Madden for insightful discussions.

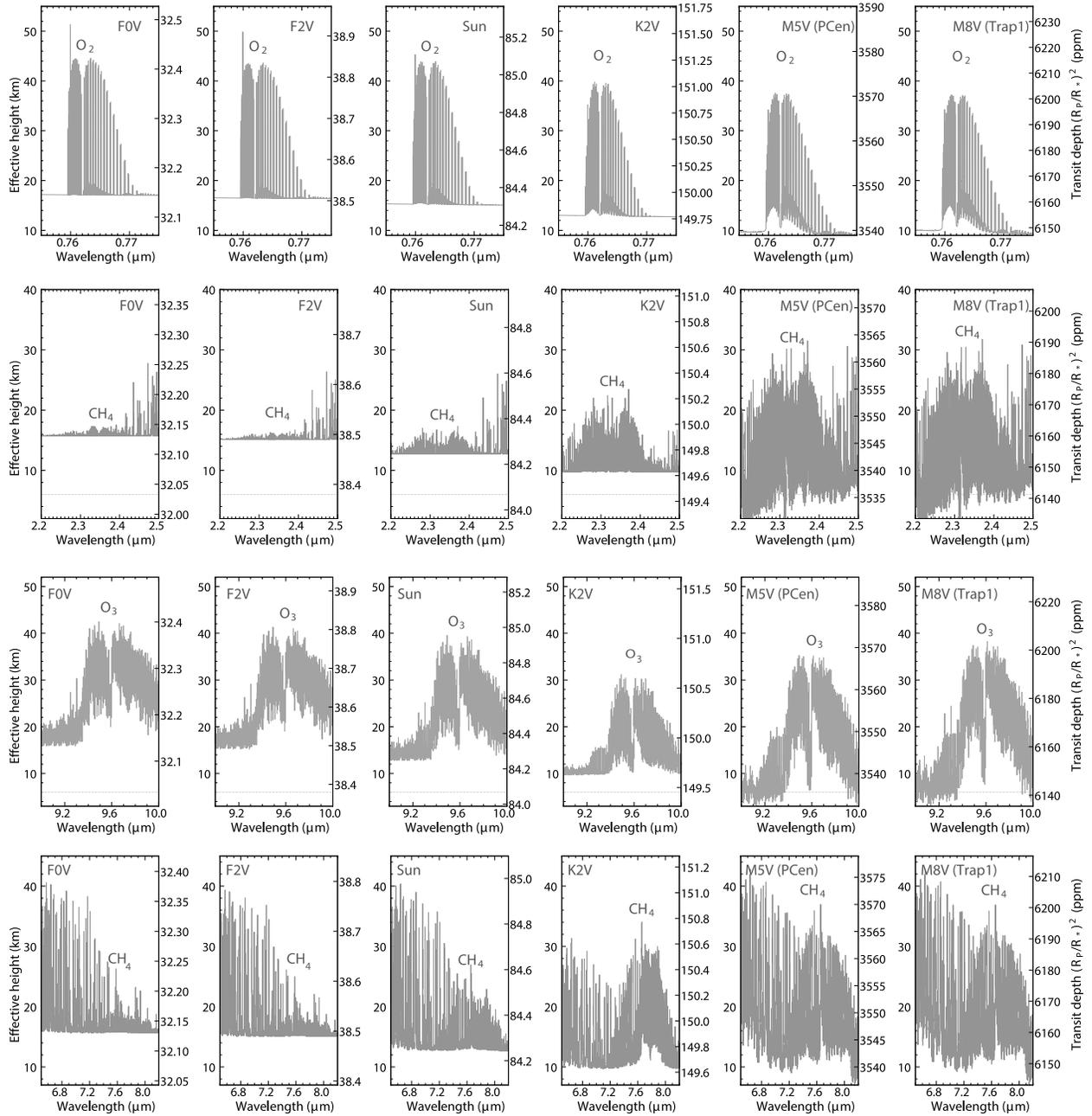

**Figure 5**: High-resolution transmission spectra for the strongest spectral features for the biosignature pairs (top) $O_2$+$CH_4$ in the Visible to Near-IR spectral range (0.4 to 3μm) and (bottom) for $O_3$+$CH_4$ in the near-IR to IR range (3 to 20μm) for Earth-like planet models orbiting 6 representative grid stars (F0,F2,Sun,K2,M5,M8). The y-axis shows effective height of the atmosphere of a planet above the ground as well as (right) transit depth, which increases by 2 orders of magnitude because of the decreasing sizes of the host star. *Data online at 10.5281/zenodo.4541545*